\documentclass[aps,showpacs,prd,twocolumn]{revtex4}
\usepackage{amsfonts}
\usepackage{amssymb}
\usepackage{amsmath}
\usepackage{color}
\usepackage[usenames,dvipsnames]{xcolor}
\usepackage{float}
\usepackage{accents}
\usepackage{graphicx}
\usepackage{epstopdf}
\usepackage{soul}
\usepackage[colorlinks=true,pdfstartview=FitV,linkcolor=blue,citecolor=blue,urlcolor=blue,breaklinks=true]{hyperref}

\begin{document}

\title{First-order vortices in a gauged $CP(2)$ model with a Chern-Simons
term}
\author{V. Almeida$^{1}$, R. Casana$^{1}$ and E. da Hora$^{2}$.}
\affiliation{$^{1}${Departamento de F\'{\i}sica, Universidade Federal do Maranh\~{a}o,}\\
65080-805, S\~{a}o Lu\'{\i}s, Maranh\~{a}o, Brazil.\\
$^{2}$Coordenadoria Interdisciplinar de Ci\^{e}ncia e Tecnologia,\\
Universidade Federal do Maranh\~{a}o, {65080-805}, S\~{a}o Lu\'{\i}s, Maranh%
\~{a}o, Brazil{.}}

\begin{abstract}
We consider a gauged $CP(2)$ theory in the presence of the Chern-Simons
action, focusing our attention on those time-independent solutions
possessing radial symmetry. In this context, we develop a coherent
first-order framework via the Bogomol'nyi prescription, from which we obtain
the corresponding energy lower-bound and the first-order equations the model
supports. We use these expressions to introduce effective BPS scenarios,
solving the resulting first-order equations by means of the
finite-difference scheme, this way attaining genuine field solutions
engendering topological configurations. We depict the new profiles,
commenting on the main properties they engender.
\end{abstract}

\maketitle

\section{Introduction}

\label{Intro}

In the context of classical theories, solitons are described as those
time-independent solutions arising within highly nonlinear models \cite{n5}.
In this sense, vortices are radially symmetric solutions coming from planar
scenarios in the presence of a gauge field.

Moreover, under very special circumstances, solitons can also be obtained
via a set of first-order differential equations (instead of the second-order
Euler-Lagrange ones), the resulting solutions minimizing the energy of the
effective system \cite{n4}.

In this sense, first-order vortices were firstly studied in the context of
the simplest Maxwell-Higgs electrodynamics \cite{n1}. Furthermore, these
solutions were verified to occur within the Chern-Simons-Higgs scenario too 
\cite{cshv}. Also, first-order vortices were recently considered in
connection with nonstandard models \cite{gaht}, the resulting solutions
being used as an attempt to explain some cosmological issues \cite{ames}.

In such a context, it is especially interesting to consider the existence of
well-behaved time-independent vortices arising from a $CP(N-1)$ scenario in
the presence of a gauge field, mainly due to the close phenomenological
relation between such theory and the four-dimensional Yang-Mills-Higgs one 
\cite{cpn-1}.

In a recent investigation, radially symmetric solutions arising from a
planar $CP\left( 2\right) $ theory endowed by the Maxwell term were
considered, the author clarifying the way these structures and correlated
results depend on the parameters of the model \cite{loginov}. In that work,
however, the vortex configurations were obtained by solving the second-order
Euler-Lagrange equations directly (the resulting solutions therefore not
saturating the Bogomol'nyi bound).

In the sequel, some of us introduced the first-order vortices inherent to
the aforementioned Maxwell $CP(2)$ theory, defining the energy lower-bound
and the corresponding first-order equations \cite{casana}. In that work, the
self-dual profiles were constructed numerically by means of the
finite-difference scheme, the resulting structures presenting the typical
topological shape.

Moreover, some of us have also studied first-order vortices within a Maxwell 
$CP(2)$ model in the presence of a nontrivial dielectric function. The point
to be raised here is that such function can be used to change the vacuum
manifold of the effective theory, from which we have used such freedom to
generate self-dual vortices engendering a nontopological profile, the
resulting Bogomol'nyi bound being not quantized anymore \cite{lima}.

We now go a little bit further by investigating a rather natural extension
of the aforecited works, i.e. the search for the first-order planar solitons
arising from a $CP(2)$ theory in the presence of the Chern-Simons action.

In order to introduce our results, the present manuscript is organized as
follows: in the next Section II, we define the gauged $CP(N-1)$ theory and
some conveniences inherent to it, focusing our attention on those
time-independent solitons possessing radial symmetry. We then develop a
coherent first-order framework by minimizing the effective energy according
the Bogomol'nyi prescription, this way obtaining general first-order
equations and the corresponding energy lower-bound, such construction being
only possible due to a differential constraint involving the potential
engendering self-duality. In the Section III, we solve the first-order
expressions in order to find genuine BPS\ solutions saturating the
Bogomol'nyi bound. We solve the corresponding first-order equations by means
of the finite-difference algorithm, from which we depict the numerical
solutions, whilst commenting the main properties they engender. We end our
work in the Section IV, presenting our final considerations and the
perspectives regarding future studies.

In this manuscript, we adopt $\eta ^{\mu \nu }=\left( +--\right) $ as the
metric signature for the flat spacetime, together with the natural units
system, for the sake of simplicity.

\section{The model \label{2}}

\label{general}

We begin our investigation by presenting the Lagrange density defining the
gauged $CP(N-1)$ model in the presence of the Chern-Simons term (with $%
\epsilon ^{012}=+1$), i.e.%
\begin{equation}
\mathcal{L}=-\frac{k}{4}\epsilon ^{\alpha \mu \nu }A_{\alpha }F_{\mu \nu
}+\left( P_{ab}D_{\mu }\phi _{b}\right) ^{\ast }P_{ac}D^{\mu }\phi
_{c}-V\left( \left\vert \phi \right\vert \right) \text{.}  \label{xxm}
\end{equation}%
Here, $F_{\mu \nu }=\partial _{\mu }A_{\nu }-\partial _{\nu }A_{\mu }$
stands for the electromagnetic field strength tensor, $D_{\mu }\phi
_{a}=\partial _{\mu }\phi _{a}-igA_{\mu }Q_{ab}\phi _{b}$ representing the
covariant derivative ($Q_{ab}$ is a real diagonal charge matrix).
Furthermore, $P_{ab}=\delta _{ab}-h^{-1}\phi _{a}\phi _{b}^{\ast }$ is a
projection operator defined conveniently. In this work, the Greek indexes
run over the space-time coordinates, the Latin ones counting the complex
fields underlying the $CP(N-1)$ sector (with $\phi _{a}^{\ast }\phi _{a}=h$).

The Euler-Lagrange equation for the gauge field is given by%
\begin{equation}
\frac{k}{2}\epsilon ^{\lambda \mu \nu }F_{\mu \nu }=J^{\lambda }\text{,}
\end{equation}%
where%
\begin{equation}
J^{\lambda }=ig\left[ \left( P_{ab}Q_{bf}\phi _{f}\right) ^{\ast
}P_{ac}D^{\lambda }\phi _{c}-\left( P_{ab}D^{\lambda }\phi _{b}\right)
^{\ast }P_{ac}Q_{cb}\phi _{b}\right]
\end{equation}%
represents the current vector. It is then instructive to write down the
Gauss law for time-independent configurations, which reads (here, $B=F_{21}$
is the magnetic field)%
\begin{equation}
kB=\rho \text{,}  \label{gl}
\end{equation}%
with%
\begin{equation}
\frac{\rho }{ig}=\left( P_{ab}D^{0}\phi _{b}\right) ^{\ast }P_{ac}Q_{cd}\phi
_{d}-P_{ab}D^{0}\phi _{b}\left( P_{ac}Q_{cd}\phi _{d}\right) ^{\ast }
\end{equation}%
and $D^{0}\phi _{b}=-igQ_{bc}\phi _{c}A^{0}$. In this sense, given that $%
A^{0}=0$ does not solve the Gauss law identically, the temporal gauge does
not hold anymore, the final structures possessing both electric and magnetic
fields.

In this work, we look for radially symmetric solutions inherent to the
gauged $CP(2)$ scenario by using the usual map%
\begin{equation}
A_{i}=-\frac{1}{gr}\epsilon ^{ij}n^{j}A(r)\text{,}  \label{xxa1}
\end{equation}%
\begin{equation}
\left( 
\begin{array}{c}
\phi _{1} \\ 
\phi _{2} \\ 
\phi _{3}%
\end{array}%
\right) =h^{\frac{1}{2}}\left( 
\begin{array}{c}
e^{im_{1}\theta }\sin \left( \alpha (r)\right) \cos \left( \beta (r)\right)
\\ 
e^{im_{2}\theta }\sin \left( \alpha (r)\right) \sin \left( \beta (r)\right)
\\ 
e^{im_{3}\theta }\cos \left( \alpha (r)\right)%
\end{array}%
\right) \text{,}  \label{xxa2}
\end{equation}%
with $m_{1}$, $m_{2}$ and $m_{3}\in \mathbb{Z}$ standing for winding
numbers, $\epsilon ^{ij}$ being the bidimensional Levi-Civita tensor (with $%
\epsilon ^{12}=+1$), $n^{j}=\left( \cos \theta ,\sin \theta \right) $
representing the unit vector. Therefore, regular solutions presenting no
divergences are obtained via those profile functions $\alpha (r)$ and $A(r)$
satisfying%
\begin{equation}
\alpha (r\rightarrow 0)\rightarrow 0\text{ \ and \ }A(r\rightarrow
0)\rightarrow 0\text{.}  \label{xxno}
\end{equation}

It is already known that, in order to support configurations with nontrivial
topology, we must fix $m_{1}=-m_{2}=m$, $m_{3}=0$ and $Q=\lambda _{3}/2$,
with $\lambda _{3}=$diag$\left( 1,-1,0\right) $ (the choice $m_{1}=m_{2}=m$%
,\ $m_{3}=0$ and $Q=\lambda _{8}/2$, with $\sqrt{3}\lambda _{8}=$diag$\left(
1,1,-2\right) $, mimicking the first one), the profile function $\beta (r)$
then holding for two constant solutions, i.e.%
\begin{equation}
\beta (r)=\beta _{1}=\frac{\pi }{4}+\frac{\pi }{2}k\text{ \ \ or \ \ }\beta
(r)=\beta _{2}=\frac{\pi }{2}k\text{,}  \label{xx1}
\end{equation}%
with $k\in Z$; for additional details, the reader is referred to the Eq. (9)
of the Ref. \cite{casana} and the discussion therein.

It is important to highlight that, from this point on, our expressions
describe the effective scenario defined by the conveniences introduced in
the previous paragraph.

We look for genuine first-order solutions saturating an energy lower-bound.
In this sense, we proceed the minimization of the overall energy, the
starting-point being the energy-momentum tensor related to the effective
scenario, i.e.%
\begin{equation}
\mathcal{T}_{\lambda \rho }=2\left( P_{ab}D_{\lambda }\phi _{b}\right)
^{\ast }P_{ac}D_{\rho }\phi _{c}-\eta _{\lambda \rho }\mathcal{L}_{ntop}%
\text{,}
\end{equation}%
where%
\begin{equation}
\mathcal{L}_{ntop}=\left( P_{ab}D_{\mu }\phi _{b}\right) ^{\ast
}P_{ac}D^{\mu }\phi _{c}-V\left( \left\vert \phi \right\vert \right)
\end{equation}%
stands for the nontopological Lagrange density, the energy density reading%
\begin{equation}
\varepsilon =\frac{k^{2}B^{2}}{g^{2}hW}+V+h\left[ \left( \frac{d\alpha }{dr}%
\right) ^{2}+\frac{W}{r^{2}}\left( \frac{A}{2}-m\right) ^{2}\right] \text{,}
\label{xxed}
\end{equation}%
where we have introduced the Gauss law (\ref{gl}). Here, we have defined the
auxiliary function%
\begin{equation}
W=W(\alpha ,\beta )=\sin ^{2}\alpha \left( 1-\sin ^{2}\alpha \cos ^{2}\left(
2\beta \right) \right) \text{.}
\end{equation}

The point to be raised is that, whether the potential is constrained to
satisfy%
\begin{equation}
\frac{2k}{g^{2}\sqrt{h}}\frac{d}{dr}\sqrt{\frac{V}{W}}=-h\sqrt{W}\frac{%
d\alpha }{dr}\text{,}  \label{xxc}
\end{equation}%
the expression for the energy density can be rewritten according the
Bogomol'nyi prescription, therefore giving rise to%
\begin{eqnarray}
\varepsilon &=&\left( \frac{kB}{g\sqrt{hW}}\mp \sqrt{V}\right) ^{2}+h\left( 
\frac{d\alpha }{dr}\mp \frac{\sqrt{W}}{r}\left( \frac{A}{2}-m\right) \right)
^{2}  \notag \\
&&\mp \frac{2k}{g^{2}\sqrt{h}}\frac{1}{r}\frac{d}{dr}\left[ \left(
A-2m\right) \sqrt{\frac{V}{W}}\right] \text{,}
\end{eqnarray}%
where we have used $B\left( r\right) =-A^{\prime }/gr$ for the magnetic
field (prime denoting derivative with respect to $r$), the resulting
first-order equations standing for%
\begin{equation}
\frac{d\alpha }{dr}=\pm \frac{\sqrt{W}}{r}\left( \frac{A}{2}-m\right) \text{,%
}  \label{xxbps1}
\end{equation}%
\begin{equation}
kB=\pm g\sqrt{hVW}\text{,}  \label{xxbps2}
\end{equation}%
the solution for $\beta (r)$ being necessarily one of those stated in (\ref%
{xx1}).

The scenario can be summarized as follows: given the potential fulfilling
the constraint (\ref{xxc}), the model (\ref{xxm}) effectively supports
radially symmetric solutions satisfying the first-order equations (\ref%
{xxbps1}) and (\ref{xxbps2}), the final configurations saturating an energy
lower-bound given by%
\begin{equation}
E_{bps}=2\pi \int r\varepsilon _{bps}dr=\mp \frac{8\pi mk}{g^{2}\sqrt{h}}%
\sqrt{\frac{V_{0}}{W_{0}}}\text{,}  \label{xxlb}
\end{equation}%
where%
\begin{equation}
\varepsilon _{bps}=\mp \frac{2k}{g^{2}\sqrt{h}}\frac{1}{r}\frac{d}{dr}\left[
\left( A-2m\right) \sqrt{\frac{V}{W}}\right]  \label{xxedbps}
\end{equation}%
stands for the energy density of the first-order structures, the upper
(lower) sign holding for negative (positive) values of $m$. Here, we have
supposed that $\left( A_{\infty }-2m\right) \sqrt{V_{\infty }/W_{\infty }}$
vanishes, with $\sqrt{V_{0}/W_{0}}$ being finite. Moreover, we have defined $%
V_{0}\equiv V\left( r\rightarrow 0\right) $, $W_{0}\equiv W\left(
r\rightarrow 0\right) $, $V_{\infty }\equiv V\left( r\rightarrow \infty
\right) $, $W_{\infty }\equiv W\left( r\rightarrow \infty \right) $ and $%
A_{\infty }\equiv A\left( r\rightarrow \infty \right) $.

It is also instructive to calculate the magnetic flux $\Phi _{B}$ the
first-order solutions support. It reads%
\begin{equation}
\Phi _{B}=2\pi \int rB\left( r\right) dr=-\frac{2\pi }{g}A_{\infty }\text{,}
\label{xxmf}
\end{equation}%
where we have used\ $B\left( r\right) =-A^{\prime }/gr$ again. We
demonstrate below that the energy lower-bound (\ref{xxlb}) can be verified
to be proportional to the magnetic flux (\ref{xxmf}), both quantities being
quantized according the winding number $m$, as expected for topological
solitons.

%%%%%%%%%%%%%%%%%%%%%%%%

\section{First-order scenarios and their numerical solutions}

We now demonstrate how the first-order framework we have developed generates
genuine radially symmetric solitons. Here, in order to present our results,
we proceed as follows: firstly, we choose a particular solution for $\beta
(r)$ coming from (\ref{xx1}), whilst solving the constraint (\ref{xxc}) for
the potential engendering self-duality. We then use such conveniences to
obtain the asymptotic boundary conditions $\alpha (r)$ and $A(r)$ must obey
in order to fulfill the finite-energy requirement, i.e. $\varepsilon
(r\rightarrow \infty )\rightarrow 0$, from which we also calculate the
energy lower-bound (\ref{xxlb}) and the magnetic flux (\ref{xxmf})
explicitly, showing that they are proportional to each other, as expected.
Finally, we solve the corresponding first-order equations numerically by
means of the finite-difference scheme, whilst commenting on the main
properties they engender.

\subsection{The $\protect\beta (r)=\protect\beta _{1}$ case}

We go further into our investigation by choosing%
\begin{equation}
\beta (r)=\beta _{1}=\frac{\pi }{4}+\frac{\pi }{2}k\text{,}  \label{v10}
\end{equation}%
from which one gets $\cos ^{2}\left( 2\beta _{1}\right) =0$, the fundamental
constraint being reduced to%
\begin{equation}
\frac{2k}{g^{2}\sqrt{h}}\frac{d}{dr}\left[ \frac{\sqrt{V}}{\sin \alpha }%
\right] =h\frac{d}{dr}\left( \cos \alpha \right) \text{,}
\end{equation}%
whose solution is%
\begin{equation}
V\left( \alpha \right) =\frac{g^{4}}{16k^{2}}h^{3}\sin ^{2}\left( 2\alpha
\right) \text{,}  \label{xxv}
\end{equation}%
i.e. the potential supporting self-duality (here, we have used $C=0$ for the
integration constant).

We now implement (\ref{v10}) and (\ref{xxv}) into (\ref{xxed}), the
resulting expression being%
\begin{eqnarray}
\varepsilon \left( r\right) &=&\frac{k^{2}B^{2}}{g^{2}h\sin ^{2}\alpha }+%
\frac{g^{4}}{16k^{2}}h^{3}\sin ^{2}\left( 2\alpha \right)  \notag \\
&&+h\left[ \left( \frac{d\alpha }{dr}\right) ^{2}+\frac{\sin ^{2}\alpha }{%
r^{2}}\left( \frac{A}{2}-m\right) ^{2}\right] \text{,}
\end{eqnarray}%
from which we attain $\varepsilon \left( r\rightarrow \infty \right)
\rightarrow 0$ by imposing%
\begin{equation}
\alpha \left( r\rightarrow \infty \right) \rightarrow \frac{\pi }{2}\text{ \
\ and \ \ }A\left( r\rightarrow \infty \right) \rightarrow 2m\text{,}
\label{xxabc}
\end{equation}%
standing for the boundary conditions the profile functions obey in the
asymptotic limit.

In view of (\ref{v10}), (\ref{xxv}) and (\ref{xxabc}), the energy
lower-bound (\ref{xxlb}) can be verified to be equal to%
\begin{equation}
E_{bps}=\mp 4\pi hm\text{,}
\end{equation}%
the magnetic flux $\Phi _{B}$ (\ref{xxmf}) standing for%
\begin{equation}
\Phi _{B}=-\frac{4\pi }{g}m\text{,}  \label{xxpvmf}
\end{equation}%
from which we get that $E_{bps}=\pm gh\Phi _{B}$, both $E_{bps}$ and $\Phi
_{B}$ being proportional to each other and quantized according the winding
number $m$, as expected. Here, we have used%
\begin{equation}
\frac{\sqrt{V_{0}}}{\sin \alpha _{0}\sqrt{1-\sin ^{2}\alpha _{0}\cos
^{2}\left( 2\beta _{1}\right) }}=\frac{g^{2}\sqrt{h}}{2k}h\text{,}
\end{equation}%
this way also verifying our previous assumption, see the discussion just
after (\ref{xxedbps}). 
\begin{figure}[tbp]
\centering\includegraphics[width=8.5cm]{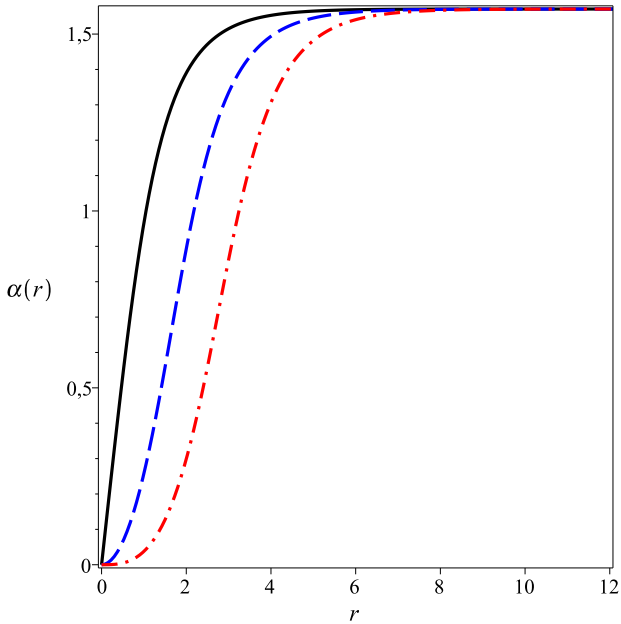}
\par
\vspace{-0.3cm}
\caption{Numerical solutions to $\protect\alpha \left( r\right) $ coming
from (\protect\ref{xbps1}) and (\protect\ref{xbps2}) in the presence of (%
\protect\ref{xxno}) and (\protect\ref{xxabc}). Here, we have fixed $h=k=1$
and $g=\protect\sqrt{2}$, varying the winding number: $m=1$ (solid black
line), $m=2$ (dashed blue line) and $m=3$ (dash-dotted red line).}
\end{figure}

The first-order equations (\ref{xxbps1}) and (\ref{xxbps2}) can be rewritten
as%
\begin{equation}
\frac{d\alpha }{dr}=\pm \frac{\sin \alpha }{r}\left( \frac{A}{2}-m\right) 
\text{,}  \label{xbps1}
\end{equation}%
\begin{equation}
\frac{1}{r}\frac{dA}{dr}=\mp \frac{g^{4}}{4k^{2}}h^{2}\sin \left( 2\alpha
\right) \sin \alpha \text{,}  \label{xbps2}
\end{equation}%
which must be solved according the boundary conditions (\ref{xxno}) and (\ref%
{xxabc}).

\subsection{The $\protect\beta (r)=\protect\beta _{2}$ case}

We now consider%
\begin{equation}
\beta (r)=\beta _{2}=\frac{\pi }{2}k\text{,}  \label{carai}
\end{equation}%
via which one gets $\cos ^{2}\left( 2\beta _{2}\right) =1$, the
corresponding constraint being%
\begin{equation}
\frac{4k}{g^{2}\sqrt{h}}\frac{d}{dr}\left[ \frac{\sqrt{V}}{\sin \left(
2\alpha \right) }\right] =\frac{h}{4}\frac{d}{dr}\left( \cos \left( 2\alpha
\right) \right) \text{,}
\end{equation}%
is solution standing for the self-dual potential, i.e.%
\begin{equation}
V\left( \alpha \right) =\frac{g^{4}}{16k^{2}}\left( \frac{h}{4}\right)
^{3}\sin ^{2}\left( 4\alpha \right) \text{,}  \label{xxv2}
\end{equation}%
where we have chosen $\mathcal{C}=0$ for the integration constant. 
\begin{figure}[tbp]
\centering\includegraphics[width=8.5cm]{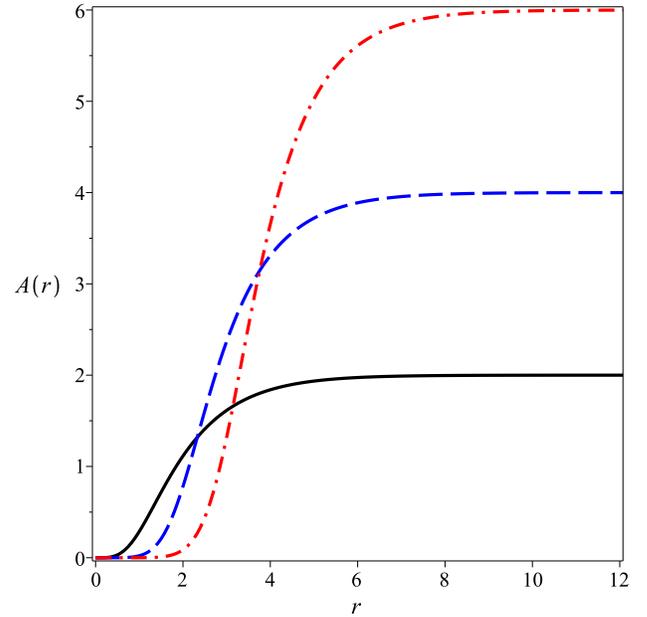}
\par
\vspace{-0.3cm}
\caption{Numerical solutions to $A\left( r\right) $. Conventions as in the
Fig. 1, the profiles being monotonic.}
\end{figure}

We proceed in the very same way as before, i.e. we use (\ref{carai}) and (%
\ref{xxv2}) into (\ref{xxed}), from which one gets the general expression%
\begin{eqnarray}
\varepsilon \left( r\right) &=&\frac{4k^{2}B^{2}}{g^{2}h\sin ^{2}\left(
2\alpha \right) }+\frac{g^{4}}{16k^{2}}\left( \frac{h}{4}\right) ^{3}\sin
^{2}\left( 4\alpha \right)  \notag \\
&&+h\left[ \left( \frac{d\alpha }{dr}\right) ^{2}+\frac{\sin ^{2}\left(
2\alpha \right) }{4r^{2}}\left( \frac{A}{2}-m\right) ^{2}\right] \text{,}
\end{eqnarray}%
the finite-energy requirement $\varepsilon \left( r\rightarrow \infty
\right) \rightarrow 0$ being attained by those profile functions fulfilling 
\begin{equation}
\alpha \left( r\rightarrow \infty \right) \rightarrow \frac{\pi }{4}\text{ \
\ and \ \ }A\left( r\rightarrow \infty \right) \rightarrow 2m\text{,}
\label{xxabc2}
\end{equation}%
i.e. the boundary conditions in the limit $r\rightarrow \infty $.

Now, due to (\ref{carai}), (\ref{xxv2}) and (\ref{xxabc2}), the energy bound
(\ref{xxlb}) reduces to 
\begin{equation}
E_{bps}=\mp \pi hm\text{,}
\end{equation}%
the magnetic flux $\Phi _{B}$ still being given by the result in (\ref%
{xxpvmf}). Therefore, one gets that $E_{bps}=\pm gh\Phi _{B}/4$, the
lower-bound being proportional to the flux of the magnetic field, both ones
being again quantized. Here, we have calculated%
\begin{equation}
\frac{\sqrt{V_{0}}}{\sin \alpha _{0}\sqrt{1-\sin ^{2}\alpha _{0}\cos
^{2}\left( 2\beta _{2}\right) }}=\frac{g^{2}\sqrt{h}}{2k}\frac{h}{4}
\end{equation}%
in order to verify our previous conjecture.

In this case, the first-order expressions (\ref{xxbps1}) and (\ref{xxbps2})
can be written in the form%
\begin{equation}
\frac{d\alpha }{dr}=\pm \frac{\sin \left( 2\alpha \right) }{2r}\left( \frac{A%
}{2}-m\right) \text{,}  \label{ybps1}
\end{equation}%
\begin{equation}
\frac{1}{r}\frac{dA}{dr}=\mp \frac{g^{4}}{4k^{2}}\left( \frac{h}{4}\right)
^{2}\sin \left( 4\alpha \right) \sin \left( 2\alpha \right) \text{,}
\label{ybps2}
\end{equation}%
which must be considered in the presence of the conditions (\ref{xxno}) and (%
\ref{xxabc2}). 
\begin{figure}[tbp]
\centering\includegraphics[width=8.5cm]{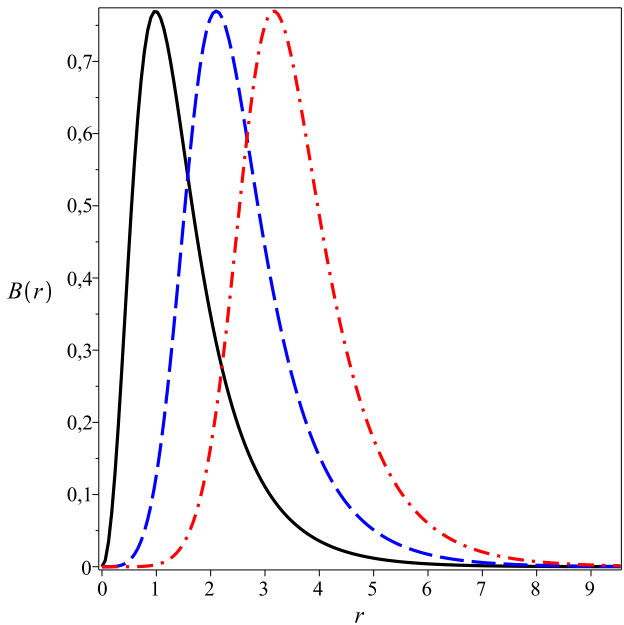}
\par
\vspace{-0.3cm}
\caption{Numerical solutions to the magnetic field $B(r)$. Conventions as in
the Fig. 1. The profiles are rings centered at $r=0$.}
\end{figure}

It is worthwhile to point out that the equations (\ref{ybps1}) and (\ref%
{ybps2}) can be obtained directly from those in (\ref{xbps1}) and (\ref%
{xbps2}) via the redefinitions $\alpha \rightarrow 2\alpha $ and $%
h\rightarrow h/4$, the energy bound and the self-dual potential behaving in
a similar way, the magnetic flux remaining the same. Therefore, given that
the two first-order scenarios introduced above are phenomenologically
equivalent, one concludes the existence of only one effective scenario. In
this sense, from now on, we focus our attention on those expressions coming
from $\beta (r)=\beta _{1}$\ only.

In what follows, we depict the results we have found by solving the
first-order equations (\ref{xbps1}) and (\ref{xbps2}) by means of the
finite-difference prescription, according the boundary conditions (\ref{xxno}%
) and (\ref{xxabc}). Here, we have considered the lower signs in the
first-order expressions (i.e. $m>0$ only), whilst choosing $h=k=1$ and $g=%
\sqrt{2}$, for the sake of simplicity. In this sense, we introduce the
solutions to the profile functions $\alpha \left( r\right) $\ and $A\left(
r\right) $, the magnetic field $B\left( r\right) $, the BPS energy density $%
\varepsilon _{bps}\left( r\right) $, the electric potential $A^{0}\left(
r\right) $ and the electric field $E\left( r\right) $ for $m=1$ (solid black
line), $m=2$ (dashed blue line) and $m=3$ (dash-dotted red line). 
\begin{figure}[tbp]
\centering\includegraphics[width=8.5cm]{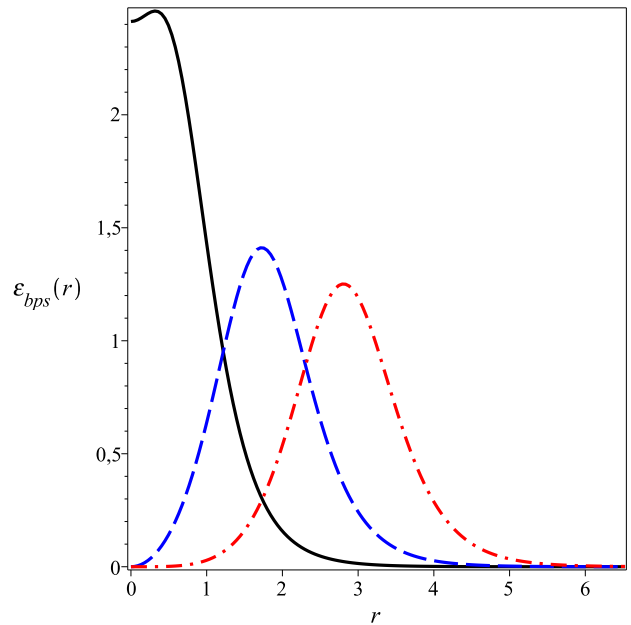}
\par
\vspace{-0.3cm}
\caption{Numerical solutions to the energy density $\protect\varepsilon %
_{bps}\left( r\right) $. Conventions as in the Fig. 1, $\protect\varepsilon %
_{bps}\left( r=0\right) $ vanishing for $m\neq 1$.}
\end{figure}

In the figures 1 and 2, we plot the numerical profiles to the functions $%
\alpha \left( r\right) $ and $A\left( r\right) $, respectively, from which
we verify the monotonic manner these fields approach the conditions (\ref%
{xxno}) and (\ref{xxabc}). In particular, we highlight the way $A\left(
r\right) $ reaches the asymptotic value $A\left( r\rightarrow \infty \right)
\rightarrow 2m$.

The Figure 3 shows the solutions to the magnetic field $B\left( r\right) $,
the resulting flux being confined on a ring centered at the origin, its
radius increasing as the winding number itself increases. It is also
interesting to note that the magnetic field vanishes asymptotically, this
way fulfilling the finite-energy requirement $\varepsilon \left(
r\rightarrow \infty \right) \rightarrow 0$.

In the Figure 4, we depict the profiles to the energy density $\varepsilon
_{bps}\left( r\right) $ inherent to the first-order configurations, these
solutions also engendering rings centered at $r=0$, their radii (amplitudes)
increasing (decreasing) as $m$ increases. Here, we point out that $%
\varepsilon _{bps}\left( r=0\right) $ vanishes for $m\neq 1$.

In the figures 5 and 6, we present the solutions to the electric potential $%
A^{0}\left( r\right) $ and to the electric field $E\left( r\right) $
inherent to it, respectively, this last one behaving in the same general way
the magnetic field does (i.e. yielding well-defined rings), both $E\left(
r=0\right) $ and $E\left( r\rightarrow \infty \right) $ vanishing
identically.

We end this Section by studying the Bogomol'nyi limit supporting
self-duality. In this sense, we proceed the linearisation of the first-order
equations (\ref{xbps1}) and (\ref{xbps2}) around the boundary values (\ref%
{xxno}) and (\ref{xxabc}), for $m>0$ (lower signs in the first-order
expressions), from we get the approximate solutions near the origin%
\begin{equation}
\alpha (r)\approx C_{0}r^{m}
\end{equation}%
and%
\begin{equation}
A(r)\approx \frac{g^{4}h^{2}C_{0}^{2}}{4k^{2}\left( m+1\right) }r^{2\left(
m+1\right) }\text{,}
\end{equation}%
the asymptotic profiles reading%
\begin{equation}
\alpha \left( r\right) \approx \frac{\pi }{2}-C_{\infty }e^{-M_{\alpha }r}
\end{equation}%
and%
\begin{equation}
A\left( r\right) \approx 2m-\frac{g^{2}h}{k}C_{\infty }re^{-M_{A}r}\text{,}
\end{equation}%
$M_{\alpha }=M_{A}=g^{2}h/2k$ being the masses of the corresponding bosons
(for $h=k=1$ and $g=\sqrt{2}$, both $M_{\alpha }$ and $M_{A}$ equal the
unity), the relation $M_{\alpha }/M_{A}=1$ defining the Bogomol'nyi limit.
Here, $C_{0}$ and $C_{\infty }$ stand for real positive integration
constants to be fixed by requiring the correct behavior at $r=0$ and $%
r\rightarrow \infty $, respectively. 
\begin{figure}[tbp]
\centering\includegraphics[width=8.5cm]{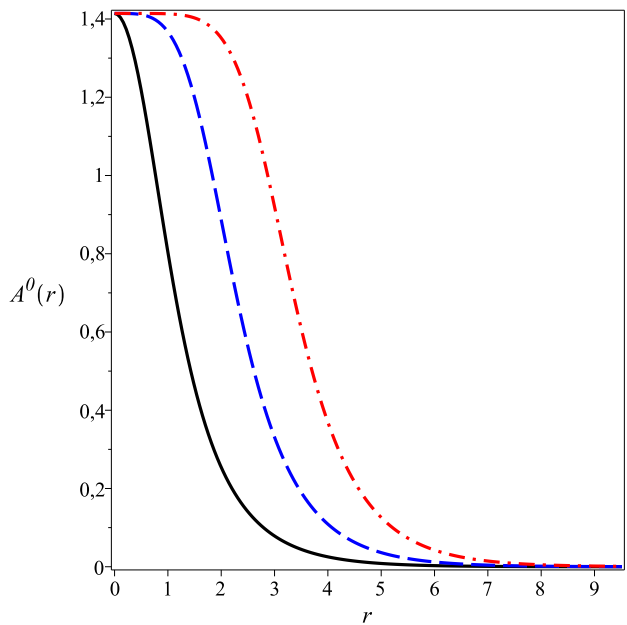}
\par
\vspace{-0.3cm}
\caption{Numerical solutions to the electric potential $A^{0}\left( r\right) 
$. Conventions as in the Fig. 1.}
\end{figure}

%%%%%%%%%%%%%%%%%%%%%%%%

\section{\textbf{Final comments and perspectives}}

We have investigated the first-order radially symmetric solutions inherent
to the $CP(2)$ model in the presence of the Chern-Simons action, from which
we have obtained regular solitons saturating a quantized energy lower-bound. 
\begin{figure}[tbp]
\centering\includegraphics[width=8.5cm]{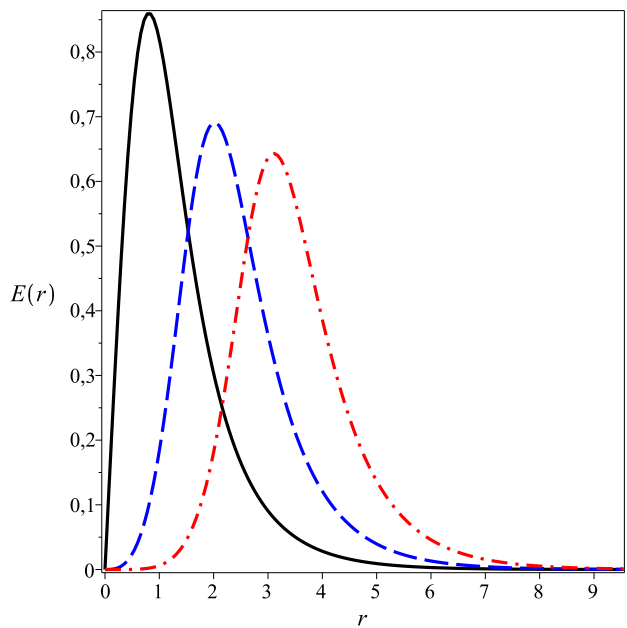}
\par
\vspace{-0.3cm}
\caption{Numerical solutions to the electric field $E\left( r\right) $.
Conventions as in the Fig. 1, both $E\left( r=0\right) $ and $E\left(
r\rightarrow \infty \right) $ vanishing.}
\end{figure}

We have introduced the overall theory and the conventions inherent to it,
focusing our attention on those time-independent configurations presenting
radial symmetry. In the sequel, we have applied the Bogomol'nyi
prescription, rewriting the expression for the effective energy in order to
introduce a well-defined lower-bound (i.e. the Bogomol'nyi bound). The point
to be raised is that such construction was only possible due to a
differential constraint involving the potential supporting self-duality.

We have considered separately the cases defined by the two different
solutions the additional profile function $\beta (r)$ supports, this way
verifying that these two contexts are\ phenomenologically equivalent,
therefore existing only one effective scenario. We have then solved the
corresponding first-order equations numerically by means of the
finite-difference algorithm, depicting the resulting profiles we have found
this way. We have pointed out the main properties the final configurations
engender, also studying the Bogomol'nyi limit explicitly.

We highlight that the results we have presented in this work only hold for
the radially symmetric structures defined by the map in (\ref{xxa1}) and (%
\ref{xxa2}), being therefore not possible to ensure that the original model
supports first-order solitons outside the radially symmetric proposal, such
question lying beyond the scope of this manuscript.

Ideas regarding future investigations include the search for the
nontopological first-order solitons coming from (\ref{xxm}) and the
development of a well-defined self-dual framework inherent to a $CP(2)$
theory in the presence of both the Maxwell and the Chern-Simons terms
simultaneously. These issues are currently under consideration, and we hope
positive results for an incoming contribution.

\begin{acknowledgments}
The authors thank CAPES, CNPq and FAPEMA (Brazilian agencies) for partial
financial support.
\end{acknowledgments}

\end{document}